\documentclass{acmsiggraph}
\usepackage{xcolor}
\usepackage{bm}
\usepackage{mathtools}
\usepackage{xcolor}
\usepackage[linesnumbered,ruled,vlined]{algorithm2e}
\DeclarePairedDelimiter\ceil{\lceil}{\rceil}

\nonstopmode

\title{Optimal Textures: Fast and Robust Texture Synthesis and Style Transfer through Optimal Transport}

\author{Eric Risser,\\Unity\:Technologies\\ \vspace{1ex}}
\pdfauthor{Eric Risser}

\TOGonlineid{}


\newlength{\h}

\setcopyright{none}

\begin{document}

 \teaser{
   \includegraphics[width=7in]{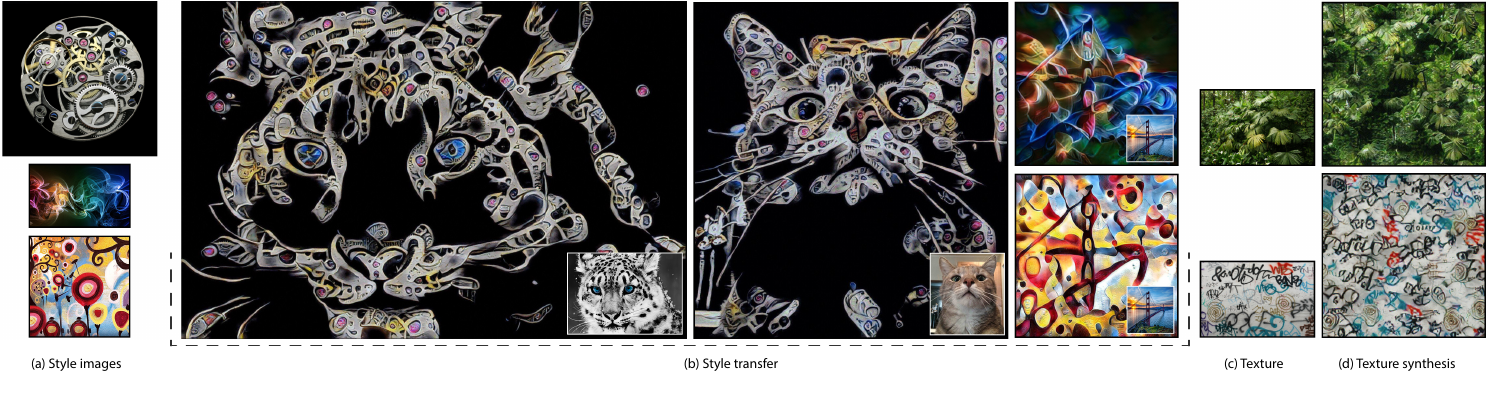}
   \caption{(a) Three style images on left followed by (b) corresponding style transfer results with content images inset in the bottom right. (c \& d) Our texture synthesis results.}
 }

\maketitle

\begin{abstract}

This paper presents a light-weight, high-quality texture synthesis algorithm that easily generalizes to other applications such as style transfer and texture mixing. We represent texture features through the deep neural activation vectors within the bottleneck layer of an auto-encoder and frame the texture synthesis problem as optimal transport between the activation values of the image being synthesized and those of an exemplar texture. To find this optimal transport mapping, we utilize an N-dimensional probability density function (PDF) transfer process that iterates over multiple random rotations of the PDF basis and matches the 1D marginal distributions across each dimension. This achieves quality and flexibility on par with expensive back-propagation based neural texture synthesis methods, but with the potential of achieving interactive rates. We demonstrate that first order statistics offer a more robust representation for texture than the second order statistics that are used today. We propose an extension of this algorithm that reduces the dimensionality of the neural feature space. We utilize a multi-scale coarse-to-fine synthesis pyramid to capture and preserve larger image features; unify color and style transfer under one framework; and further augment this system with a novel masking scheme that re-samples and re-weights the feature distribution for user-guided texture painting and targeted style transfer.

\end{abstract}

%
%


%
%


\keywordlist

\conceptlist

\printcopyright

\section{Introduction}

Methods for both representing and synthesizing textures have been explored broadly. Recently, focus has gravitated towards utilizing neural networks, both as a way to represent texture features as well as a mechanism for performing synthesis. The seminal work by Gatys et al. \shortcite{gatys2015texture} shows that the correlation of features extracted by a deep neural network (i.e. the Gram matrix) can function as a fully parametric summary of texture characteristics. Since then, hundreds of follow-up papers have better/faster ways of performing neural texture synthesis through minimizing the distance between correlation matrices or other approximations of the textures feature distribution. This paper deviates from this trend and proposes a statistically motivated formulation of the Texture Synthesis problem as one of robust feature transformation through optimal transport, with contributions over the state-of-the-art in neural texture synthesis in two areas: \textbf{performance} and \textbf{generalization}.

By \textbf{performance}, we refer to both the visual quality as well as the speed of the algorithm, as this is a trade-off. Our approach achieves superior results with a small computational budget by deviating from prior art in neural texture synthesis. Typically such approaches deeply entangle the two problems, representation and generation, solving them both in tandem through either back-propagation optimization or fully feed-forward methods. In contrast, we propose a light-weight optimization process, an N-Dimensional probability density function transform operating directly on the deep neural features themselves, within the bottleneck layer of an auto-encoder. This achieves the quality and flexibility of expensive back-propagation based methods but within a fast feed-forward auto-encoder framework that does not require custom training. We further accelerate the N-Dimensional PDF transform through dimension reduction.

By \textbf{generalization}, we refer to our statistically-motivated approach being a general algorithm that envelops other texture synthesis problems such as Style Transfer, Inverse Texture Synthesis and Texture Mixing. These classically difficult problems have historically required significant modifications to popular texture synthesis algorithms, or justified their own custom tailored approach. We show that our statistically motivated approach more directly represents the native texture synthesis problem and can solve these special cases either directly or with minor modifications. In addition, the generation process can be directly influenced through re-sampling the feature distribution based on user-drawn masks. This allows for guided, controllable synthesis with only minor changes. 

Overall, the speed, quality and generalization of our approach leads to a neural network-based solution for texture synthesis problems that is viable for use in industry.

\textbf{Our contributions include:}
\begin{enumerate}
\item     A fast, high quality neural texture synthesis method based on robust feature matching of first order statistics. We present the first optimization based neural texture synthesis method that executes directly in feature space, not requiring back-propagation training.  
\item     An acceleration strategy making this approach interactive, even for high resolution images.
\item     Extensions to several special case problems such as Style Transfer and Texture Mixing. 
\item     A unified statistical model for style and color transfer using a single algorithm.
\item     A novel user control scheme, based on feature re-sampling through guide maps.
\end{enumerate}

\section{Related work}

Methods for both representing and synthesizing textures have been explored broadly over the last decades. Heeger and Bergen \shortcite{heeger1995pyramid} represented texture using only first-order feature statistics gathered through convolution of the image with a filter bank and utilized an optimization process to transform a noise image into one that statistically matches an exemplar. Portilla and Simoncelli \shortcite{portilla2000parametric} expanded this concept with more sophisticated filters and an emphasis on the joint Nth-order statistics of the filter responses, averaged across the image into a parametric model. 

\textbf{Patch-based methods} represent texture as a collection of overlapping image patches and the various corresponding synthesis methods  attempt to re-arrange the configuration of the patches ~\cite{efros1999texture,wei2000fast,hertzmann2001image,lefebvre2005parallel,lefebvre2006appearance,Wei08InverseTextureSynthesis,barnes2009patchmatch} and blend their overlapping regions so that the resulting image shares similar patch statistics as the exemplar~\cite{kwatra2005texture,darabi2012image}.

\textbf{Deep Learning-based algorithms} have achieved state of the art results on classically difficult special cases of the texture synthesis problem, predominantly Style Transfer. The seminal work on neural texture synthesis and style transfer ~\cite{gatys2015texture,gatys2016image}, introduced deep learning to the field, significantly advancing the quality of textures synthesized from a parametric model. This work builds upon an image synthesis strategy first used for visualizing the training process within a CNN ~\cite{Mahendran14NeuralInversion} and later extended by DeepDream to produce artistic work ~\cite{DeepDream}. Inspired by Portilla and Simoncelli \shortcite{portilla2000parametric}, a collection of Gram matrices gathered from several key layers of a neural network are cumulatively used as the parametric model for texture, where transforming an image to mimic the texture of another is achieved through minimizing the distance between each image's respective set of Gram matrices. Since neural texture synthesis introduced the concept, it has become common practice to numerically measure the visual similarity of two textures as the distance between their corresponding averaged co-occurrence matrices. Several techniques have been developed to improve synthesis quality. An inherent instability of the Gram matrix based parametric model is highlighted and the loss function is supplemented with an additional histogram matching term  ~\cite{Risser17}, similar to the first order statistics matching approach first presented by Heeger and Bergen. They also introduced a coarse-to-fine multi-scale pyramid approach for the synthesis process which yielded both speed and quality improvements. Many other contemporary extensions to the basic Gatys approach were proposed to extend its functionality for related image synthesis tasks such as regular pattern synthesis ~\cite{Sendik17DeepCorrelations} and Texture Painting ~\cite{Gatys17Painting}.
 
A major drawback of the Gatys et al. method is the high cost of utilizing back-propagation training as a general purpose optimizer for texture synthesis. To address this, several feed-forward network training schemes have been explored to approximate the optimization process, formulating the problem as one of learning texture synthesis as an image-to-image translation problem ~\cite{johnson2016,ulyanov2016texture}. While fast, these inference methods are comparatively weaker with respect to visual quality and they require training one network for one or a small number of styles. Thus, much of the research in this area has been focused on improving visual quality ~\cite{Wang17FastMultiModalStyleTransfer,Li17DiversifiedTextureSynthesis} and arbitrary texture support ~\cite{Chen16VGGAutoEncoder,Li2017WCT}. 

\textbf{The first truly universal style transfer} method that did not require custom training for each style was introduced by Chen and Schmidt \shortcite{chen2016fast} who present an auto-encoder strategy that mimics the original back-propagation strategy of Gatys. They used pre-trained VGG as the encoder and trained a VGG inversion network as the decoder. Their generation method was based on earlier non-parametric patch-based synthesis. This strategy was expanded upon by Li et al. \shortcite{Li17DiversifiedTextureSynthesis} introducing decoders after each pooling layer of VGG and a deep-to-shallow iterative synthesis strategy, more closely mimicking the original Gatys approach that matches a set of layers for each pooling size.

\subsection{Theoretical Motivation}

The goal of texture synthesis is, given an exemplar image, to construe a generative process that can synthesize arbitrarily many new unique images that are statistically indistinguishable from the exemplar. Textures are stationary by definition, therefore texture can be modeled as a finite set of statistical measurements taken over the spatial extent of a theoretically infinite image. Any sub-infinite image with the same statistical measurements is therefore considered the same texture. Modeling texture in such a way conveniently provides many mathematical tools to analytically measure the similarity between two textures. 

The study of texture synthesis can be broadly summarized as having two goals: ~\textbf{(1)} finding better representations for texture that more directly model the key feature statistics and ~\textbf{(2)} finding better generative processes for synthesizing new images that match to a set of exemplar feature statistics. These two goals are symbiotic in nature and should be designed to work in tandem, reinforcing each others strengths. Evidence suggests this is not happening in modern optimization-based neural texture synthesis approaches. This point is highlighted by two papers that achieve near comparable results to Gatys et al. while simplifying opposing aspects of the original paper. Li et al. \shortcite{Li17DiversifiedTextureSynthesis} continue using pre-trained VGG as their feature space representations but replace the expensive optimization-based generative process with the Whitening Coloring Transform, an approximation that can be solved in closed form. In contrast, Ustyuzhaninov et al. \shortcite{Ustyuzhaninov16ShallowRandomNetworks} found that Gatys's generative process applied to a single layer network with random weights can perform texture synthesis on par with the full approach. This shows that neither deep neural networks nor the fact that they were trained to learn a feature space that mimics human vision are a contributing factor to the success of the original method, rather, the powerful general purpose LBFGS optimizer is capable of brute forcing a texture synthesis solution despite a weak representation. Surprisingly, one paper shows that a strong feature representation can achieve good results despite a basic generation approach while the other paper shows that robust generation approach makes the choice of representation irrelevant.

~\textbf{We make two observations:} (1) Back-propagation based optimization, while powerful, is superfluous and therefore inefficient when using pre-trained VGG as the feature representation. (2) It is surprising that the original approach by Gatys does not achieve greatly superior results to Li et al. or Ustyuzhaninov et al. despite having both a strong feature representation as well as a strong generation process. This warrants a deeper look at the factor that all three papers have in common, the use of second-order summary statistics and the case for parametric synthesis in general.

 While parametric synthesis is theoretically interesting, if the goal is to achieve high quality and user controllable synthesis, it has many disadvantages and no clear advantages. We observe that a parametric model is a summary representation of an underlying feature distribution, so minimizing the distance between the feature distributions directly will also achieve a minimization of the parametric model. The stationary nature of textures motivated early researchers to discard spatial information by opting to use summary statistics as their model for texture. However, summary statistics are not necessary for discarding spatial information, but they do discard feature information. Therefore, they are a weaker representation than modeling the feature statistics directly and likely the fundamental limitation in Gatys and follow-up literature.

\subsection{The case for first order statistics}

 This observation is reinforced by the analysis of the instabilities inherent with the Gram model (and parametric models in general)  ~\cite{Risser17}, with the proposed solution being an additional direct matching of first order statistics during the optimization process, achieving superior synthesis quality over the original approach by Gatys. They use a set of 1D histogram matches across channels in a neural network layer. This idea shares similarities with much earlier works ~\cite{heeger1995pyramid} who posed their generative process as one of matching first order feature statistics through a set of 1D axis-aligned histogram swaps on feature distributions. Inspired by the success in the color transfer literature with matching first order statistics in a robust manner through an N-Dimensional Probability Density Function transformation ~\cite{Pitie07ColorTransfer}, Rabin et al. \shortcite{PDF_Texture} combine the two methods, applying N-Dimensional Probability Density Function transformation to texture synthesis using a wavelet pyramid representation. While this approach focused on the problem of texture mixtures, it reinforces our theoretical motivation for using optimal transport as a texture generation process.  

\begin{figure}[hbt!]
    \centering
	\setlength{\h}{1.4in}
	\includegraphics[height=\h]{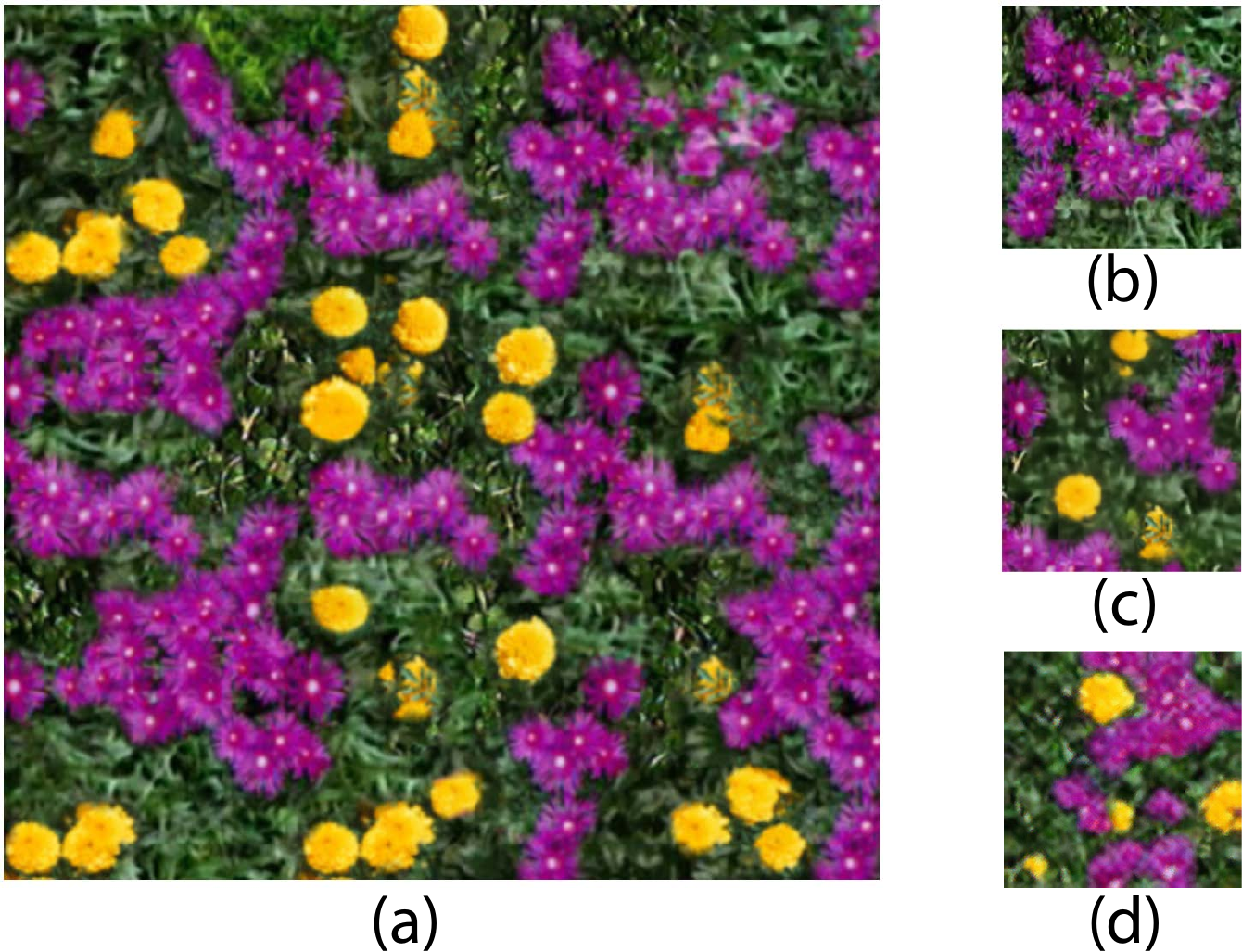}
	\caption{Comparison against Inverse Texture Synthesis. (a) The input exemplar. (b) Synthesis result using single-direction nearest neighbor search, gets stuck in local minimum and does not represent the entire image. (c) Bi-directional nearest neighbor search matches global image statistics. (c) Our optimal transport approach achieves similar results.}
	\label{fig:inverseTextureSynthesis}
\end{figure}

In parallel to the exploration of filter-based methods that match first order statistics, many patch-based methods were developed to also match first order statistics. Patch-based methods are designed as Markov random fields (MRF), where the interaction of overlapping patches within a neighborhood characterizes the statistical model for texture. Patch-based methods utilize a nearest neighbor search strategy to directly match patches from the synthesized texture with ground truth patches in the exemplar. The goal is to update pixel values or blend patches so that the resulting synthesis patch more closely mimics the exemplar patch. In this regard, the generative process is just matching first order statistics directly. It's interesting to note that early patch-based methods only matched a forward term where every pixel in the synthesis image finds its nearest neighbor in the exemplar image. Inverse Texture Synthesis ~\cite{Wei08InverseTextureSynthesis} highlighted this as a weakness in the approach and augmented it with an inverse term to also minimize the error between each pixel in the exemplar and its nearest neighbor in the synthesis image. The forward and inverse terms work together as an optimal transport strategy for matching exact feature statistics between two Probability Density Functions, as illustrated in figure ~\ref{fig:inverseTextureSynthesis}. We make the second observation that these linear filter and patch-based threads of research independently converge on a \textbf{similar conclusion:} Optimal transport of first order feature distributions is a superior texture generation process. It logically follows that this would be true as well for the non-linear-filter based methods of modern neural network texture synthesis.

\section{Proposed Algorithm}

Previous non-linear filter-based neural network methods can be grouped into two broad categories: back-propagation optimization and feed-forward. Each category has been characterized by specific shortcomings with respect to either speed, quality or the ability to generalize to multiple textures. The feed-forward literature largely identifies optimization in general as the cause of poor performance, not the back-propagation method specifically. Due to the belief that optimization is inherently slow, the previous literature focuses exclusively on approximating the final result of optimization through closed-form solutions and custom network training. We challenge this belief and introduce a hybrid approach, a method offering the benefits of both the optimization and inference categories while avoiding their respective shortcomings. The key idea is to retain a robust optimization process that matches feature statistics, but move the process from image space deep into the networks feature space. Therefore, we remove the neural network transform from the expensive optimization loop and instead transform the deep neural network activation values directly. This allows us to ~\textbf{(1)} avoid the considerable overhead of inferring the neural network in each optimization pass and ~\textbf{(2)} frame the problem in a more straightforward way, making larger gains per iteration significantly reducing the overall number of steps.
 
Our proposed algorithm combines the best ideas from previously unrelated streams of research, pairing a relatively strong and computationally efficient representation with a relatively robust and computationally efficient generative process. It is by no means an exhaustive study of all combinations of representations and generative processes, therefore, we do not make the claim that this is the theoretically optimal solution. Rather, this is a practical system that outperforms the current state of the art, performing well across the key trade-offs: visual quality, speed and generalization. This system is motivated by key insights gained through studying the underlying theory of texture synthesis.

Our algorithm mimics the back-propagation texture optimization process through an optimal transport-based feature transformation within the bottleneck layer of a series of multi-scale auto-encoder loops, similar to the strategy presented in Universal Style Transfer via Feature Transforms ~\cite{Li17DiversifiedTextureSynthesis}. In this framework VGG-19 ~\cite{Simonyan14c} pre-trained for computer vision is used as the encoder. A collection of layers are chosen from the VGG network to represent image features at different sizes and degrees of complexity. For each of these target layers a decoder network symmetric to VGG-19 (up to that target layer) is trained to invert feature space back into the original image. For consistency we adopt the same target layers and decoder networks used in the previous work.

Formally, for texture synthesis the goal is to take an input source texture $S$ and synthesize a unique but visually similar output texture $O$. This is achieved by passing both $S$ and $O$ through VGG-19 and gathering the resulting $N$ feature maps for the activations at a target layer $l$. This is denoted as $S_l$ and $O_l$ where $l$ denotes one of the following layers: Relu5\_1, Relu4\_1, Relu3\_1, Relu2\_1 and Relu1\_1.

~\subsection{Optimal Transport}

Given a pair of $N$-dimensional feature distributions $S_l$ and $O_l$, optimal transport is used to modify the activation values for $O_l$ so that the first order statistics match those of $S_l$ before decoding back into image space. We use the sliced histogram matching approach provided in the related color transfer ~\cite{Pitie07ColorTransfer} and texture mixing ~\cite{PDF_Texture} literature. By "slice" we refer to taking a random N-Dimensional unit vector, building a new N-Dimensional basis orthogonal to that vector and projecting the PDF onto that new basis. This is equivalent to a random rotation of the PDF. Following the neural texture synthesis literature, this feature transformation can be seen as a generalization of the previously proposed histogram distance ~\cite{Risser17} to operate on a random orthogonal basis of the $N$-dimensional space. By performing this process in an iterative loop across many random basis, the process robustly matches feature interdependence between the dimensions. This robust feature matching achieves the same goal as the Gram/Covariance, making them unnecessary and leaving only the sliced histogram distance as a term to be minimized. This has the advantage of simplifying the original feature distance by Risser, removing the need for two distance functions that are difficult to jointly minimize. 

\begin{figure}[hbt!]
    \centering
	\setlength{\h}{1.6in}
	\includegraphics[height=\h]{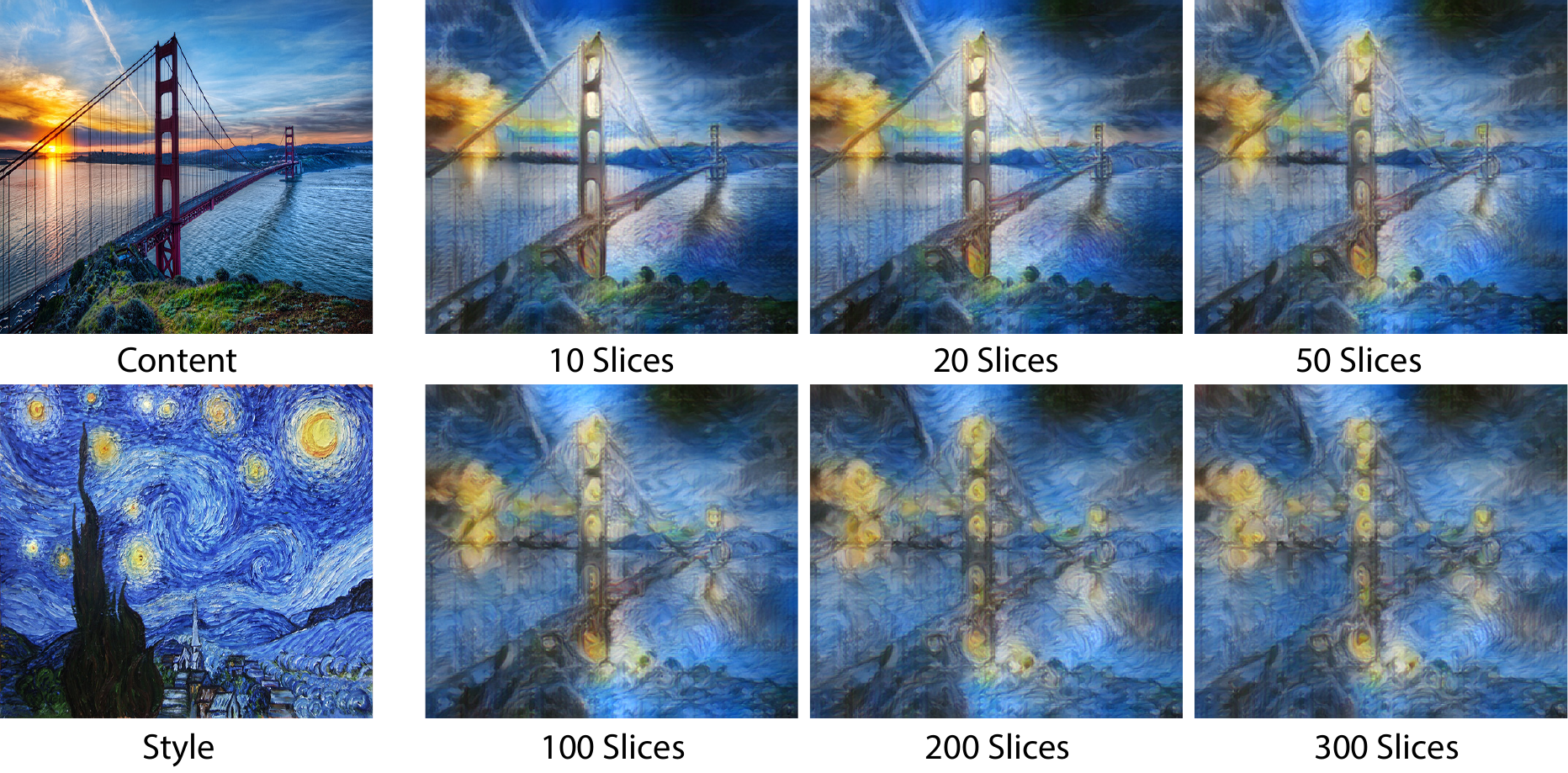}
	\caption{Increasing the number of histogram slices correlates visually to an increase in style similarity. The larger swirls and higher complexity star features become prominent as the number of slices are increased. These images were processed through the Relu5\_1 autoencoder only, starting from the content image.}
	\label{fig:featureOptimization}
\end{figure}

Iterative reduction of feature distance through sliced histogram matching within an auto-encoder offers a best-of-both-worlds solution, as it maintains the robust feature transformation of a back-propagation method while achieving the efficiency of feed-forward methods. As seen in figure ~\ref{fig:featureOptimization}, the accuracy of texture features being transferred is directly proportional to the number of slices matched. 

~\subsection{Full Algorithm}

The one advantage of a back-propagation based image reconstruction method over our series of chained auto-encoders is back-propagation's ability to optimize the image being generated towards multiple PDF targets during each iteration. Within the VGG network, we match PDFs at layers: Relu5\_1, Relu4\_1, Relu3\_1, Relu2\_1 and Relu1\_1 in that ordering. Because we cycle through $image->encoder->transform->decoder->image$ for each layer, each layer is optimally transported in isolation from the others and optimal matches at coarse layers can drift from their ideal state as the process moves to shallow layers. This problem can be mitigated with an additional global loop that runs the entire process multiple times. To keep the algorithm fast, the number of random slices can be reduced in each pass so that the total number of slices are maintained. This achieves the same effect as back-propagation of keeping all layers optimized jointly. In practice we find that only a small number of global iterations are necessary to achieve good alignment between the layers (3-6 loops depending on speed/quality trade off).

\begin{algorithm}
\SetKwInput{KwInput}{Input}                
\SetKwInput{KwOutput}{Output}              
\DontPrintSemicolon
  
  \KwInput{texture image $S$}
  \KwOutput{output image $O$}

  \SetKwFunction{FMain}{Main}
  \SetKwFunction{FSum}{OT}
  \SetKwFunction{FSub}{MatchSlice}
 
  \SetKwProg{Fn}{Function}{:}{\KwRet}
  \Fn{\FMain}{
        O = noise\;
        globalPasses = 5\;
        \For{globalLooper \textbf{=} 0 \textbf{to} globalPasses \textbf{step} 1}
        {
            \For{layer \textbf{=} 5 \textbf{to} 1 \textbf{step} 1}
            {
           	    S\_layer = VGG[layer](S)\;
           		O\_layer = VGG[layer](O)\;
           		O\_layer = OT(O\_layer, S\_layer, globalPasses)\;
           		O = VGG\_Decoder[layer](O\_layer)\;
       		}
        }
        \KwRet O\;
  }

  \SetKwProg{Fn}{Function}{:}{}
  \Fn{\FSum{$O\_layer$, $S\_layer$, $passes$}}{
        N = getTensorChannels(O\_layer)\;
        $sliceCount = N \div passes$\;
        \For{slice \textbf{=} 0 \textbf{to} sliceCount \textbf{step} 1}
        {
            basis = randomBasis(N)\;
            rotated-S\_layer = Project(basis, S\_layer)\;
            rotated-O\_layer = Project(basis, P\_layer)\;
            rotated-O\_layer = MatchSlice(rotated-O\_layer, rotated-S\_layer)\;
            O\_layer = DeProject(basis, rotated-O\_layer)\;
   		}
        \KwRet O\_layer\;
  }

  \SetKwProg{Fn}{Function}{:}{}
  \Fn{\FSub{$O\_layer$, $S\_layer$}}{
        bins = 128\;
        N = getTensorChannels(O\_layer)\;
        \For{dimLooper \textbf{=} 0 \textbf{to} N \textbf{step} 1}
        {   O\_dim = O\_layer[:, :, dimLooper]\;
            S\_dim = S\_layer[:, :, dimLooper]\;
            O\_dim = MatchHistogram(bins, O\_dim, S\_dim)\;
            O\_layer[:, :, dimLooper] = O\_dim\;
        }
        \KwRet O\_layer\;
  }

\end{algorithm}

\subsection{Optimal Transport on Principal Components}

Feature space resulting from the VGG transformation to deeper layers of the network becomes increasingly sparse. This implies that the representation for texture exists in a lower dimensional subspace of the one produced by VGG. We exploit this characteristic to accelerate the algorithm by performing optimal transport on the subspace identified through Principle Component Analysis (PCA). This extension to the algorithm does not require any modifications to the optimal transport algorithm presented. PCA is carried out for the texture/style features at the bottleneck layer of the auto-encoder and all network features are projected onto this basis. We choose the top N basis vectors of highest variance that cumulatively account for 90\% of total variance.

\begin{figure}[hbt]
    \centering
	\setlength{\h}{1in}
	\includegraphics[height=\h]{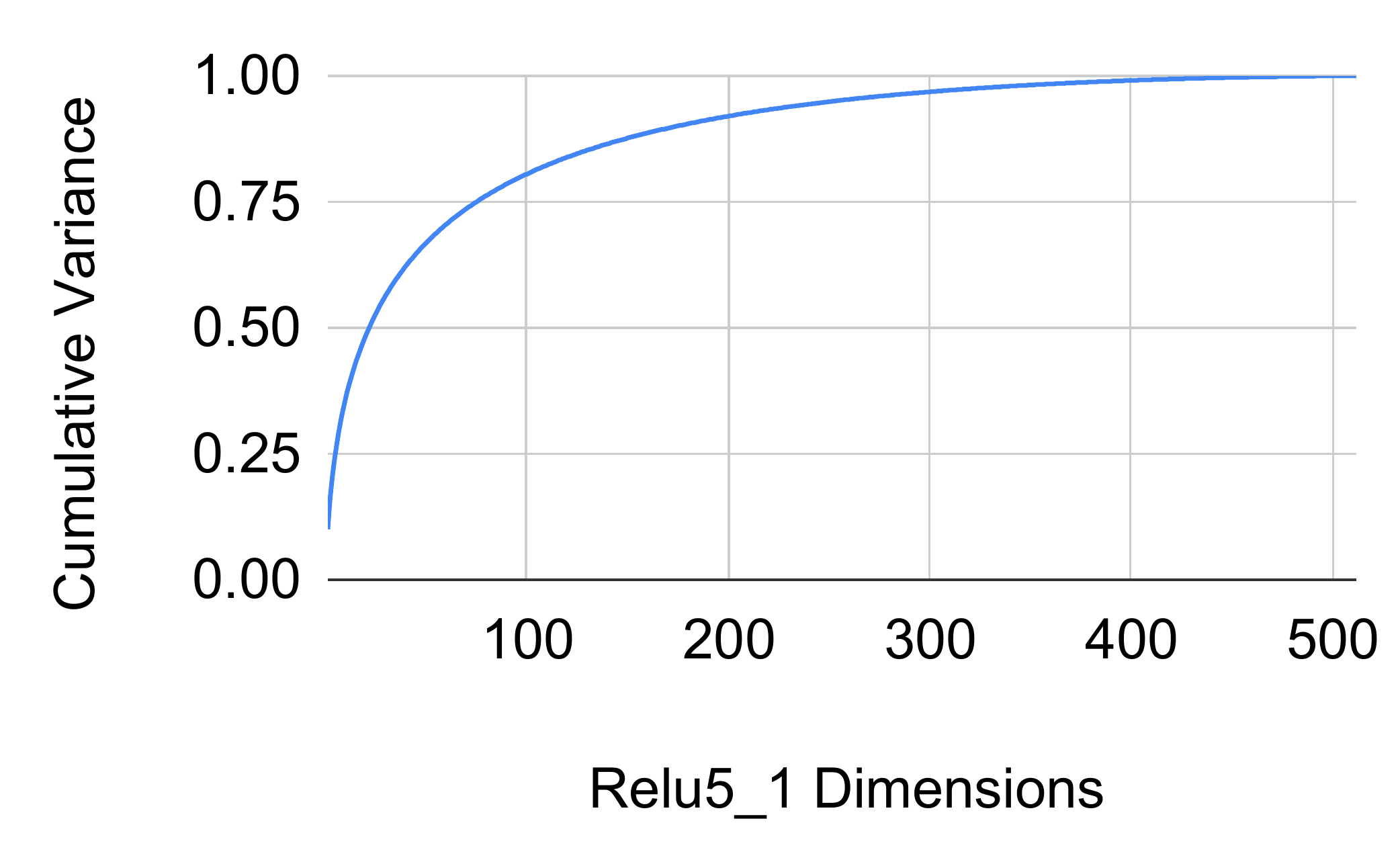}
	\includegraphics[height=\h]{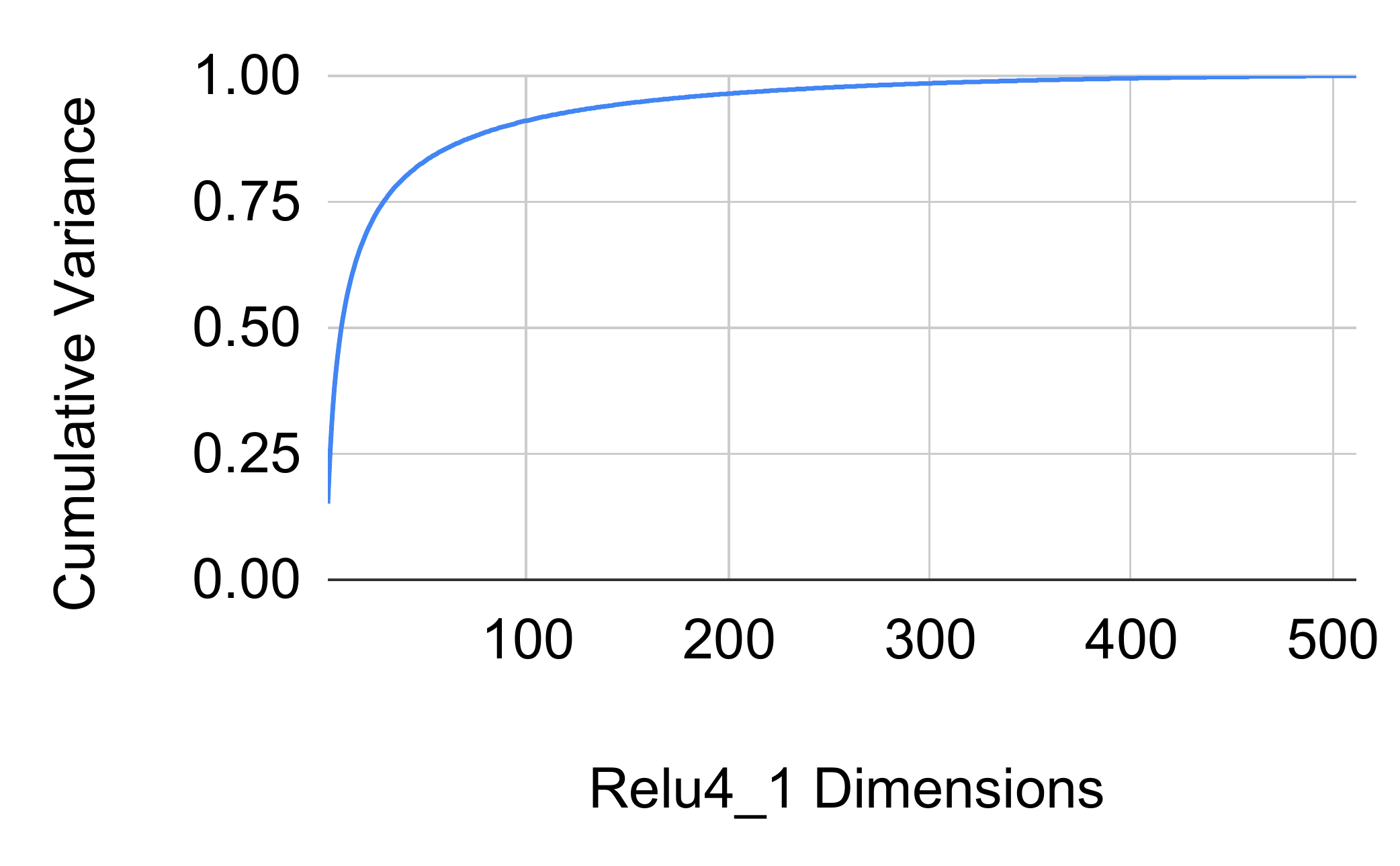}
	\includegraphics[height=\h]{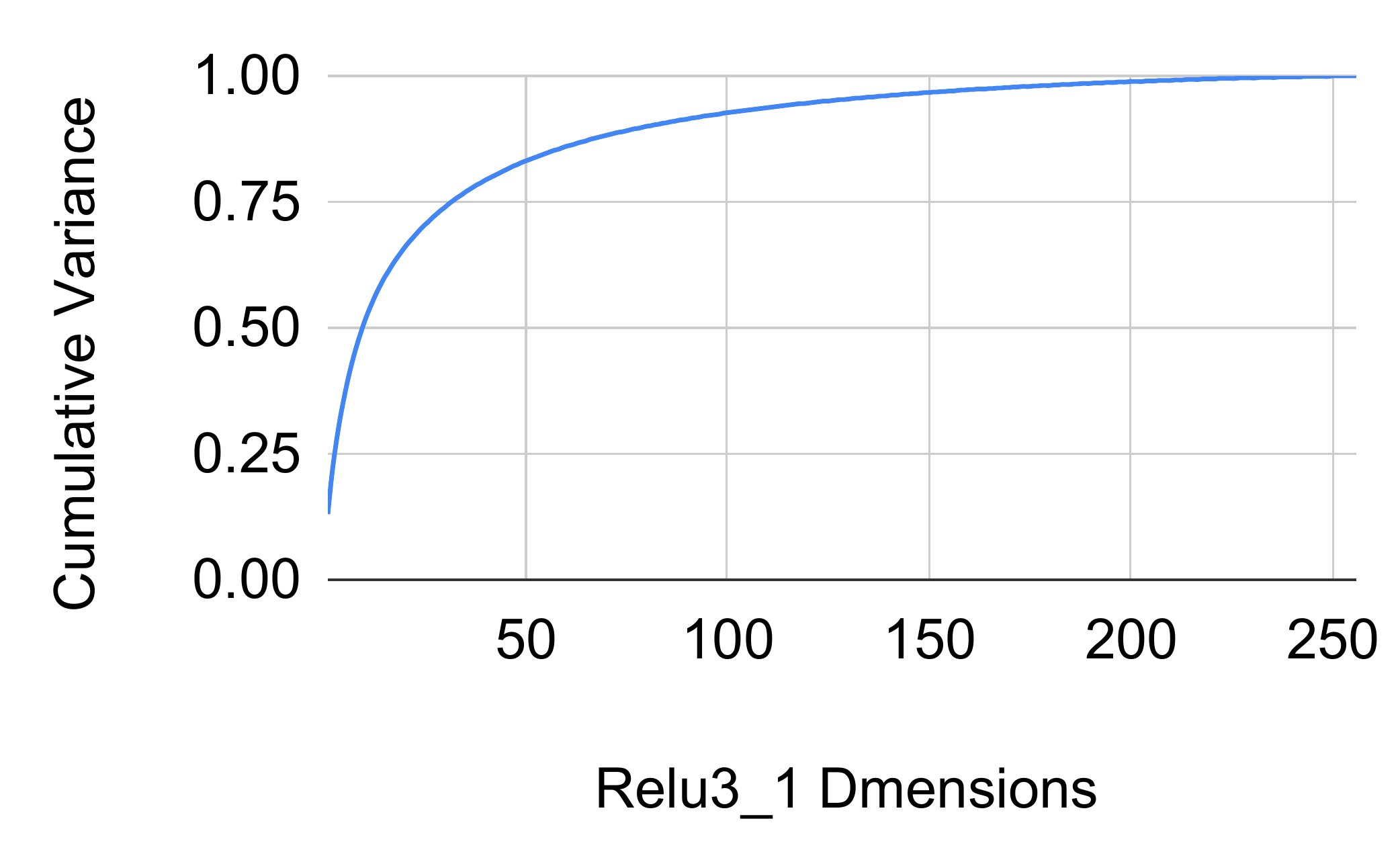}
	\includegraphics[height=\h]{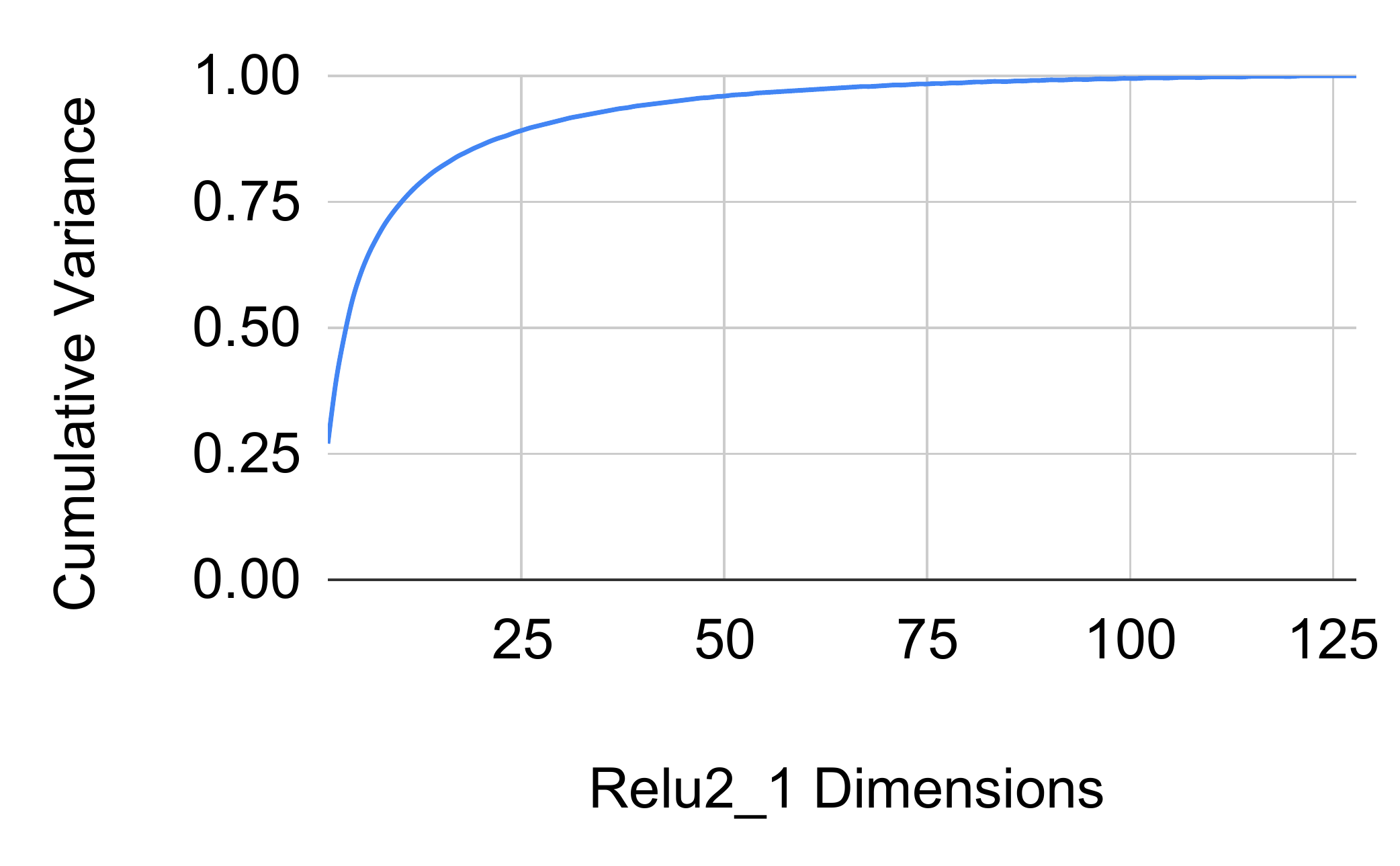}
	\includegraphics[height=\h]{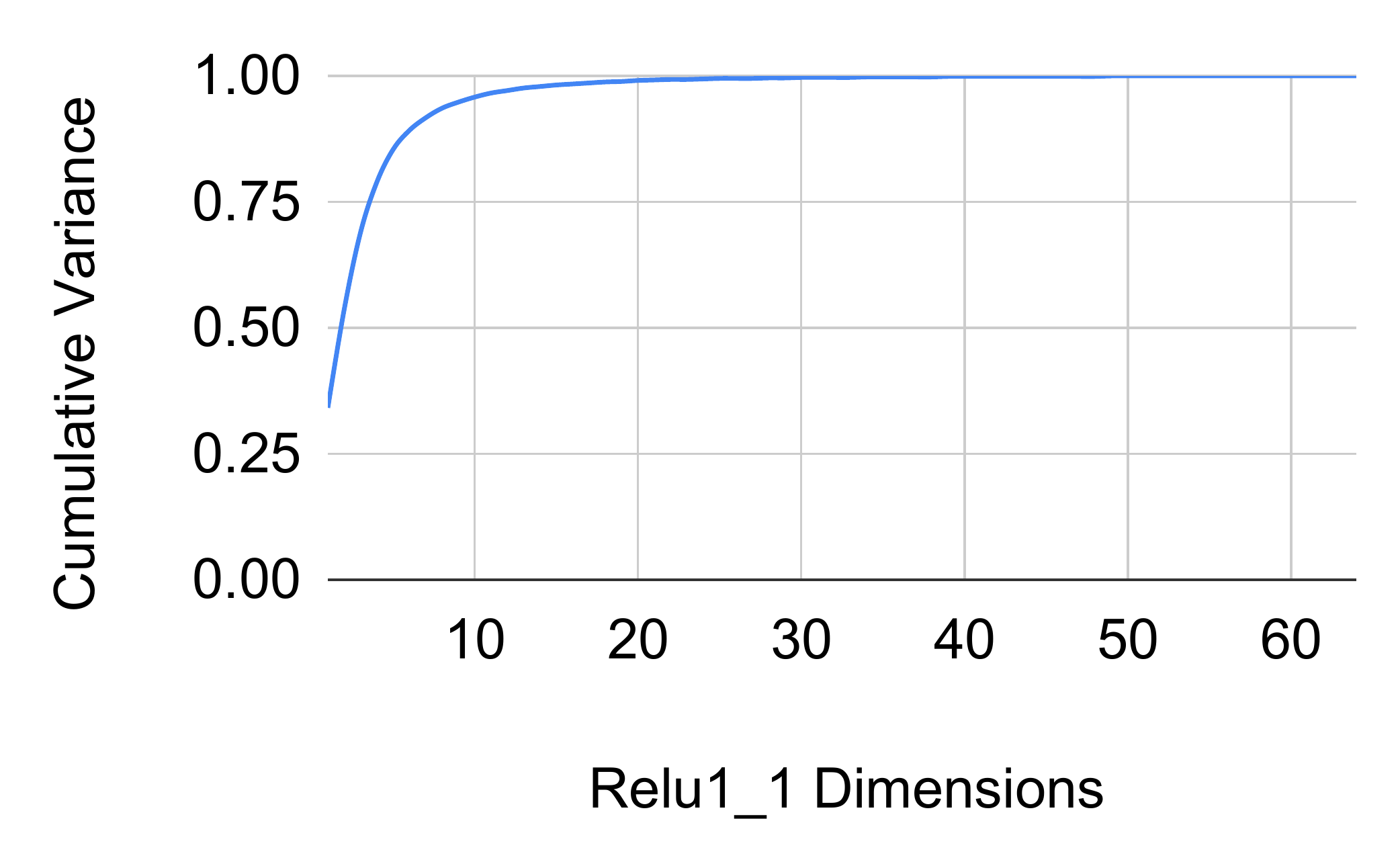}
	\caption{Cumulative variance of each layers PCA basis, averaged over 100 random textures.}
	\label{fig:PCA}
\end{figure}

We have analyzed the cumulative variance of each layers PCA basis, averaged over 100 random textures as shown in figure ~\ref{fig:PCA}. We have confirmed that our feature space can be effectively represented by a much lower dimensional subspace. 

\begin{figure}[hbt!]
	\centering
	\includegraphics[height=2.2in]{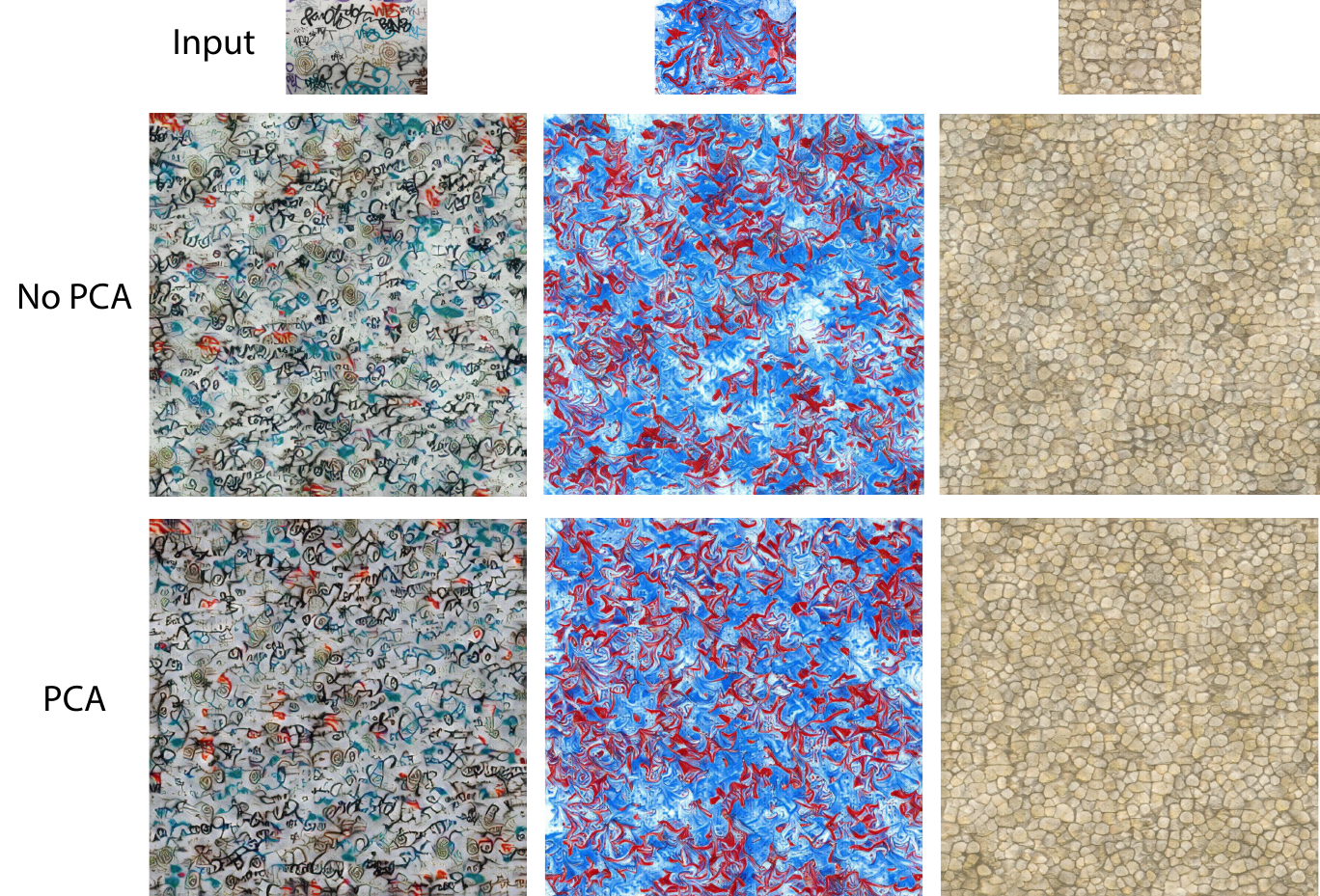}
	\caption{We observe that texture can be matched through a lower dimensional subspace of VGG.} 
	\label{fig:PCAComparison}
\end{figure}

This observation is further reinforced empirically through visual quality of the results as illustrated in figure ~\ref{fig:PCAComparison} where a side-by-side comparison is given using the full VGG layer vs. the principle components that account for the top 90\% of variance within the space. We observe marginal perceptual difference when introducing PCA, only a small bias towards generating globally homogeneous textures. This reinforces our belief that our feature representation for texture exists in a lower dimensional subspace of VGG.  

\begin{figure}[hbt!]
	\centering
	\includegraphics[height=3.2in]{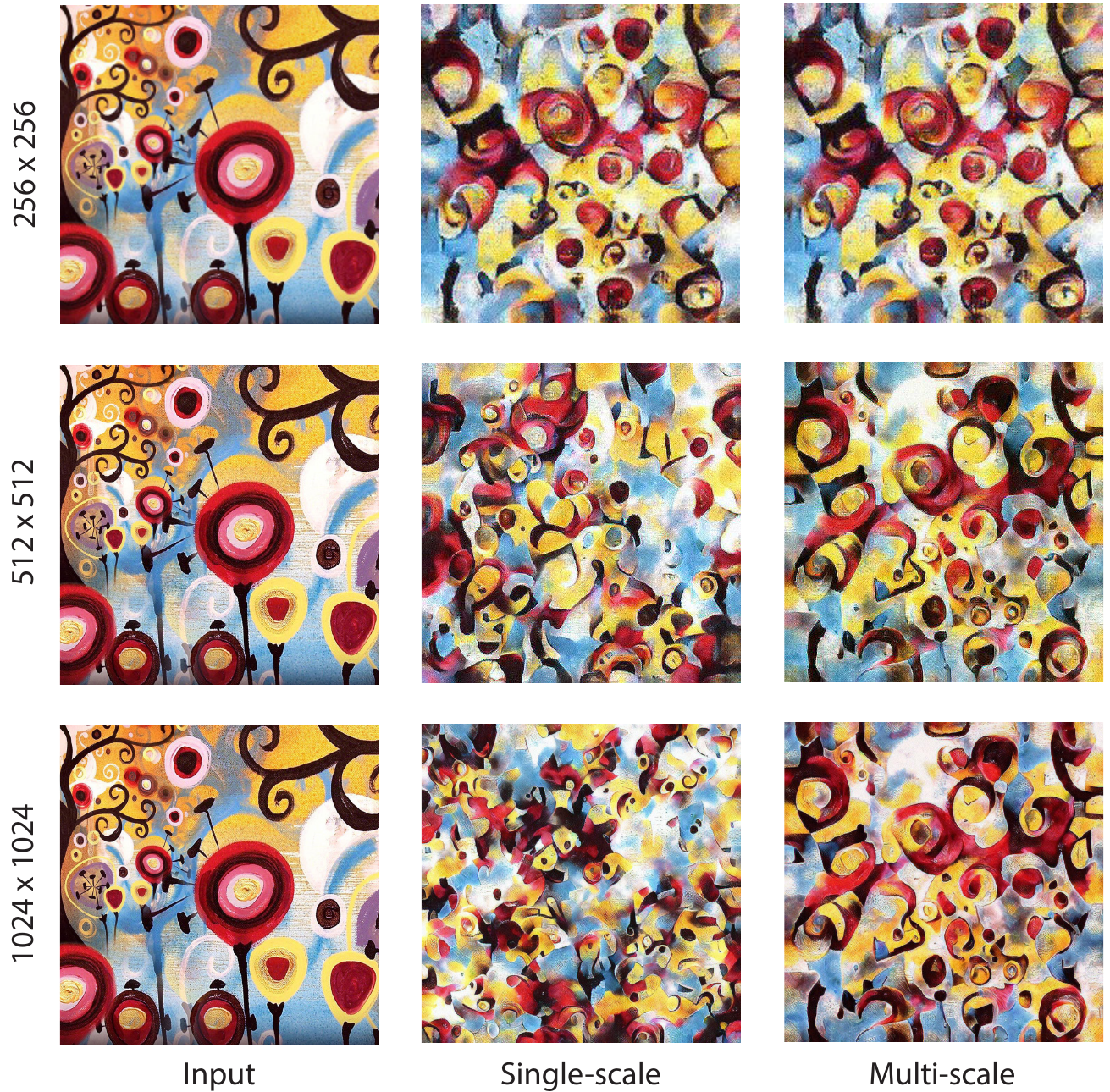}
	\caption{Top Row: input, mutli-scale and single-scale synthesis are shown at 256x256 resolution. Being the base resolution, multi-scale only operates on a single level and therefore produces the same result as single-scale. Middle Row: Increasing the resolution to 512x512 shows how multi-scale and single-scale strategies differ. Multi-scale maintains and refines the global pattern of the top row while the single-scale algorithm produces a different global distribution and loses the larger circular features present in the input. Third Row: Increasing the resolution of all images to 1024x1024 highlights that the mutli-scale approach is able to consistently refine the previous level. The single-scale approach can only represent the smallest of image features at this resolution and fails to reproduce much of the input texture.} 
	\label{fig:multiscaleComparison}
\end{figure}

\subsection{Multiresolution Synthesis}
\label{sec:pyramid}

For texture synthesis, style transfer and mixing, we have found results are generally improved by a coarse-to-fine synthesis using the same image pyramids strategy introduced by Risser et al. \shortcite{Risser17}. Given both the exemplar images and desired synthesis image resolutions, we build a pyramid by successively dividing the image widths and heights by a ratio of two until any image in the set falls below 256 pixels in either dimension. This ensures that the receptive field has sufficient coverage at the coarsest pyramid level in order to represent large structures in the feature space. The synthesis results of one pyramid level is up-scaled to the resolution of the next pyramid level using a bicubic filter and further refined through repeating the full algorithm. The Coarse-to-fine image-pyramid synthesis strategy makes it possible to synthesize large and complex texture or style features for images of a resolution necessary for real-world use as illustrated in figure ~\ref{fig:multiscaleComparison}. We use pyramids for all results in this paper unless otherwise indicated.

\begin{figure*}
	\centering
	\includegraphics[width=17.5cm]{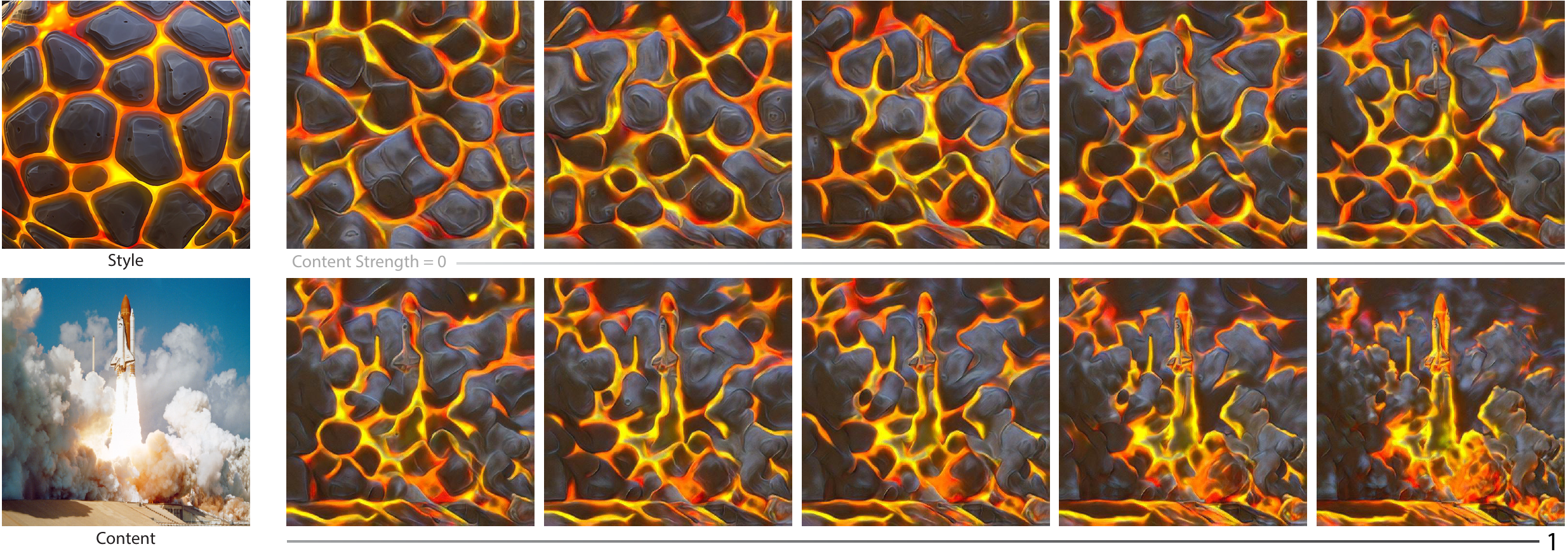}
	\caption{This illustration highlights the effects of content strength weighting on the style transfer process, showing a selection of values ranging betwene 0 and 1 for the same content and style image pair.} 
	\label{fig:contentStrength}
\end{figure*}

\begin{figure}[hbt!]
	\centering
	\setlength{\h}{2.2in}
	\includegraphics[height=\h]{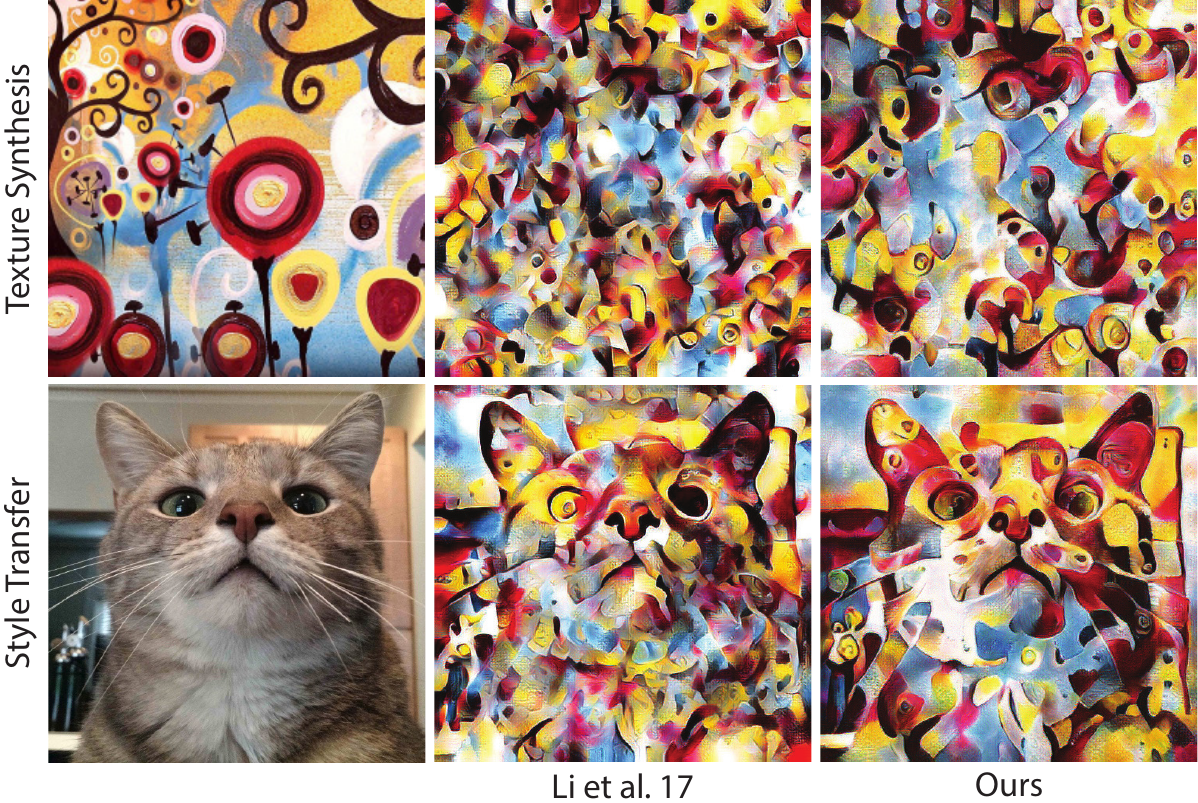}
	\caption{Note: these images are 512x512 and do not use the multiresolution image-pyramid synthesis strategy for either result. Our optimal transport algorithm can be directly compared against the WCT of Li et al. For both texture synthesis and style transfer we see that optimal transport is superior at reproducing the texture/style features while also suffering from fewer feature blending/smearing artifacts. In particular, note for Style Transfer that Optimal Transport does not only do a superior job at reproducing the style, but it also outperforms WCT at preserving the content features as well. This shows that our method is superior at finding a new unique PDF that better represents both images, rather than simply playing "tug-of-war" against the two.} 
	\label{fig:WCTComparison}
\end{figure}

\section{Extensions to Other Applications}

Optimal transport offers an intuitive and principled framework for generalizing texture synthesis to the highly-related problems of style transfer and texture mixing. Within an optimal transport framework the problem of texture synthesis is one of synthesizing an output image $O$ locally that exhibits the same global first order feature statistics of some exemplar source texture image $S$ across the range of all meaningful feature sizes. 

\subsection{Style Transfer}
\label{sec:style}

Style transfer expands upon the texture synthesis problem statement by introducing a second exemplar image, a "content image" $C$ which is also matched during synthesis, but weighted so the synthesis PDF favors the content image at coarser features while favoring the style/texture image $S$ at the finer features. In addition, the content PDF is matched in a non-local way, where pixel coordinates targets specific locations in feature space.

Optimal Transport through sliced histogram matching is uniquely well suited for high quality style transfer within a fast feed-forward approach due to its iterative nature. Before optimization, we first align $C$ by subtracting out its mean and adding the mean of $S$. During optimization, after each MatchSlice operation and subsequent de-projection, a MatchContent operation is run that updates $O\_layer$ using the equation $O\_layer = O\_layer + (C\_layer - O\_layer)\times contentStrength$. Where $contentStrength$ is a user controllable scalar that determines the degree of influence that the content image has on the final output $O$ as shown in figure \ref{fig:contentStrength}. When performing Style Transfer we apply a content matching operation only at layers Relu5\_1, Relu4\_1 and Relu3\_1, where $contentStrength$ is divided by 2 for for Relu4\_1 and divided by 4 at Relu3\_1.

Our optimal transport algorithm is an optimization process. Because we pair a content match after each iterative slice of the style match algorithm. This results in a style transfer algorithm where the content and style features optimize together, rather than the "tug-of-war" behavior seen by the WCT approach. This is reflected in figure ~\ref{fig:WCTComparison} where both the style and content feature are more prevalent in our approach over that of ~\cite{Li17DiversifiedTextureSynthesis}. This subtle but important distinction is why our approach is able to achieve style transfer results akin to back-propagation methods, but using a fast feed-forward approach.   

\subsection{Color}

\begin{figure*}
	\centering
	\includegraphics[width=17.5cm]{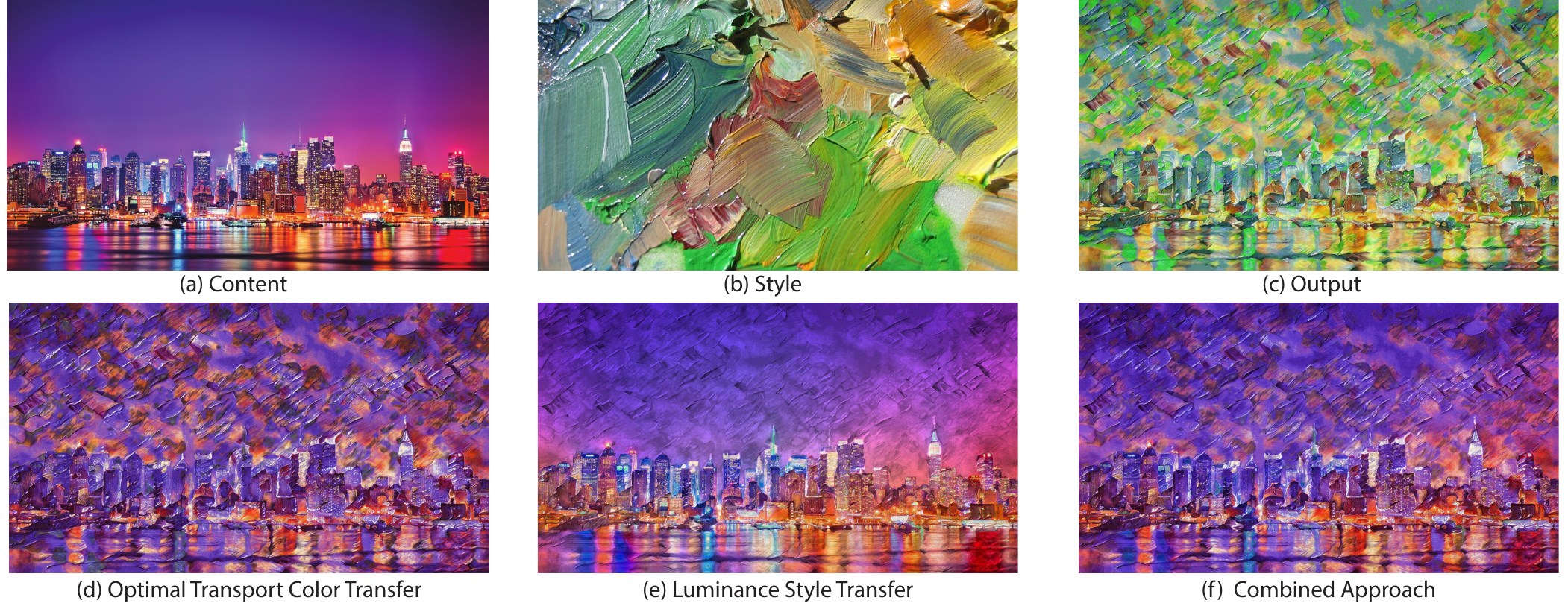}
	\caption{Images (a) and (b) show content and style images while image (c) shows our result without color transfer. Image (d) shows a global color transfer of content images onto our result. Notice that the local colors of objects in the content image are not transferred, only the global appearance. Image (e) shows the luminance result, where the effect of style transfer is softened and the brush strokes do not align with the colors. Image (f) is our combined approach where image (e) is used to locally anchor color values during the global optimal transport process.} 
	\label{fig:Color}
\end{figure*}

The original neural style transfer algorithm entangles color within the feature representation and therefore transfer of the style image's colors is intrinsic to the algorithm. Gatys highlights this as a potential shortcoming of the original method ~\cite{Gatys16Color} and explores two ideas for retaining the content image's colors. These two methods include: (1) a basic histogram match of second order color statistics, explored as a separate process from the style transfer algorithm. (2) Luminance-only style transfer of greyscale images, where the original colors of the content image are directly copied into the final result. Gatys explores multiple tweaks to these algorithms and compares the various strengths and weaknesses of each option, concluding that both solutions are viable in some situations but neither solution is strictly correct. They close by suggesting that future work should unify the two statistical models of color and CNN activations into a single framework. 

We present optimal transport as this unified framework, combining color and feature under one model, manipulated with a single algorithm. Control over color is achieved by using the three-dimensional color values directly as a final probability density function that sits on top of the multi-scale auto-encoder stack. Relative to the second order histogram matching of the previous work, our optimal transport-based color transfer achieves a more accurate mapping of the content image colors, as studied in the color transfer literature ~\cite{Pitie07ColorTransfer}. In addition, we combine the strengths of both direct color transfer and luminance based style transfer. Direct color transfer is a global operation that does not preserve the local colors. Luminance based style transfer weakens the overall style transfer effect while dependencies between luminance and the color channels are lost in the output. This is particularly apparent for styles with prominent brushstrokes as colors do not align to the stroke pattern. We propose a combined method that utilizes both strategies within the unified framework, overcoming each of their respective limitations. 

The first step in our combined approach is to reproduce the luminance based style transfer proposed by Gatys. Starting from the content image $C$ and the final output $O$ of our style transfer process, we convert both from RGB to HSL color space. We combine the hue and saturation (HS) components from the content image and the light (L) component from our style transfer result. This is illustrated in figure \ref{fig:Color} where $C$ is shown in (a), $O$ is shown in (c) and (e) is produced by taking the HS components from (a) and the L component from (c). We convert the final result back into RGB space, which is used for the remainder of the color transfer process.
Next we perform our full optimal transport algorithm that we have presented for style transfer but we use the three-channel RGB values of each image directly rather than the activation values produced by VGG. When performing our optimal transport algorithm for color transfer, we replace our $C$ image with the "luminance style transfer" that we just created. We update $S$ with the original input content image, because this image contains the color properties that we want transferred. Our optimization process thus robustly transfers the global content image color statistics while also anchoring specific colors to local image regions. Again referring to figure \ref{fig:Color}, this process is illustrated in (d) which shows our optimal transport algorithm starting from (c) and using (a) as the source image $S$ and no content image $C$. While this robustly captures the color statistics of (a), it fails to do so in a manner that retains the original location of the colors. By introducing (e) into the optimal transport process as a content image $C$, the algorithm is able to robustly transfer the colors of (a) while anchoring those colors to their desired locations as shown in (f). 

\subsection{Texture Mixing}
\label{sec:mixing}

\begin{figure*}[ht]
	\centering
	\includegraphics[width=17.5cm]{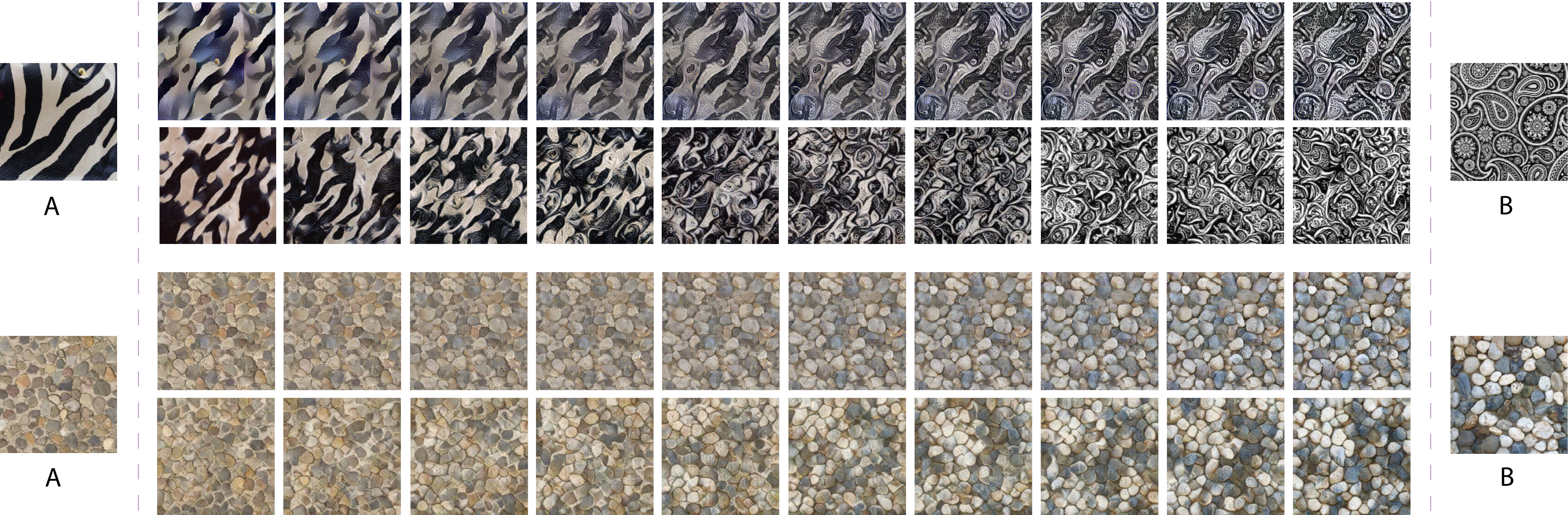}
	\caption{Comparisons of texture mixing against Wang et al. 2018 (rows 1 and 3) and our optimal transport scheme (rows 2 and 4). } 
	\label{fig:mixing}
\end{figure*}

The goal of texture mixing is to interpolate and blend the features of two or more textures. This can be used to create novel hybrid textures or to create the complex transitions between different textures within a painting application. A naive interpolation of multiple distinct textures at the pixel level will lead to ghosting, seam and blurring artifacts and will not produce novel interpolations of the texture features. The topic of texture mixing has been studied through the main bodies of texture synthesis research: non parametric ~\cite{darabi2012image,diamanti2015synthesis}, parametric: ~\cite{heeger1995pyramid,portilla2000parametric} and recent neural approaches, some of which building from the recent parametric neural texture synthesis algorithm ~\cite{Li17DiversifiedTextureSynthesis,Zi-Ming18Mixing}. Recently, methods for texture mixing have adopted an adversarial strategy to learn a custom latent space for texture features ~\cite{yu2019texture}. While these techniques offer many viable methods for texture mixing, we believe there is still an opportunity to solve the mixing problem in a way that is: fast, works on a broad range of textures, is simple to implement, does not require custom training and generates high quality results.

We propose optimal transport within a deep neural feature space as a viable method for achieving all these goals. We pose this strategy as a modernization of the earlier work on optimal transport for texture mixing ~\cite{PDF_Texture} with the main deviation from the earlier approach being the use of a trained CNN as the transformation function between image and feature space. The earlier work employed a steerable wavelet pyramid as their feature representation, which was considered state-of-the-art at the time, but was not able to achieve visually pleasing results.

Mixing two textures $A$ and $B$ can be achieved through interpolation of their first order statistics, yielding a "mixed" feature distribution that is used for $S$ in the synthesis algorithm. Producing this mixed feature distribution can be achieved by first computing the optimal transport mapping from $A$ to $B$, $A_B$. A naive solution would be to directly interpolate the values where $S = A \times (1 - i) + A_B \times i$ where $i$ is the interpolation value between 0 and 1. While this approach achieves satisfactory results in many scenarios, the optimal transport mapping is an approximate operation and can lead to a small degree of deviation from the original distribution. This results in an algorithm where synthesis reproduction quality is superior for texture $A$ as it is only mapped once during synthesis while texture $B$ is first mapped during interpolation and the result is then remapped a second time during synthesis, leading to compounding error.

To achieve uniform synthesis quality, we introduce a second optimal transport mapping from $B$ to $A$, $B_A$ as well as a "mixing mask" that contains a random 0-1 interpolation value for each pixel, following a uniform distribution across the image. We generate a mixed $S$ using the following equation: 

\begin{equation*} \label{eq:1}
    mix = \ceil*{mixingMask - i}
\end{equation*}
  
\begin{equation*} \label{eq:2}
    S = (A \times (1 - i) + A_B \times i) \times mix + (B_A \times (1 - i) + B \times i) \times (1 - mix)
\end{equation*}

Mixing through optimal transport achieves state-of-the-art results without the need for custom training. We compare our results against two recent neural network based texture synthesis mixing techniques. Figure \ref{fig:mixing} compares our approach against the method presented by \cite{Zi-Ming18Mixing} that extends the earlier Gram matrix-based neural texture synthesis algorithm for the problem of mixing. The results are comparable with ours as both methods produce new hybrid features that share characteristics of both exemplar texture $A$ and $B$ and both methods are able to smoothly interpolate this feature hybridization. Our method in some cases is able to achieve a more accurate reproduction of the input textures and runs orders of magnitude faster. Compared against \cite{Li17DiversifiedTextureSynthesis}, our approach is able to achieve a superior hybridization at the feature level and does not exhibit the spatial "tug-of-war" appearance between incompatible features.  

\begin{figure}[hbt!]
	\centering
	\setlength{\h}{3in}
	\includegraphics[height=\h]{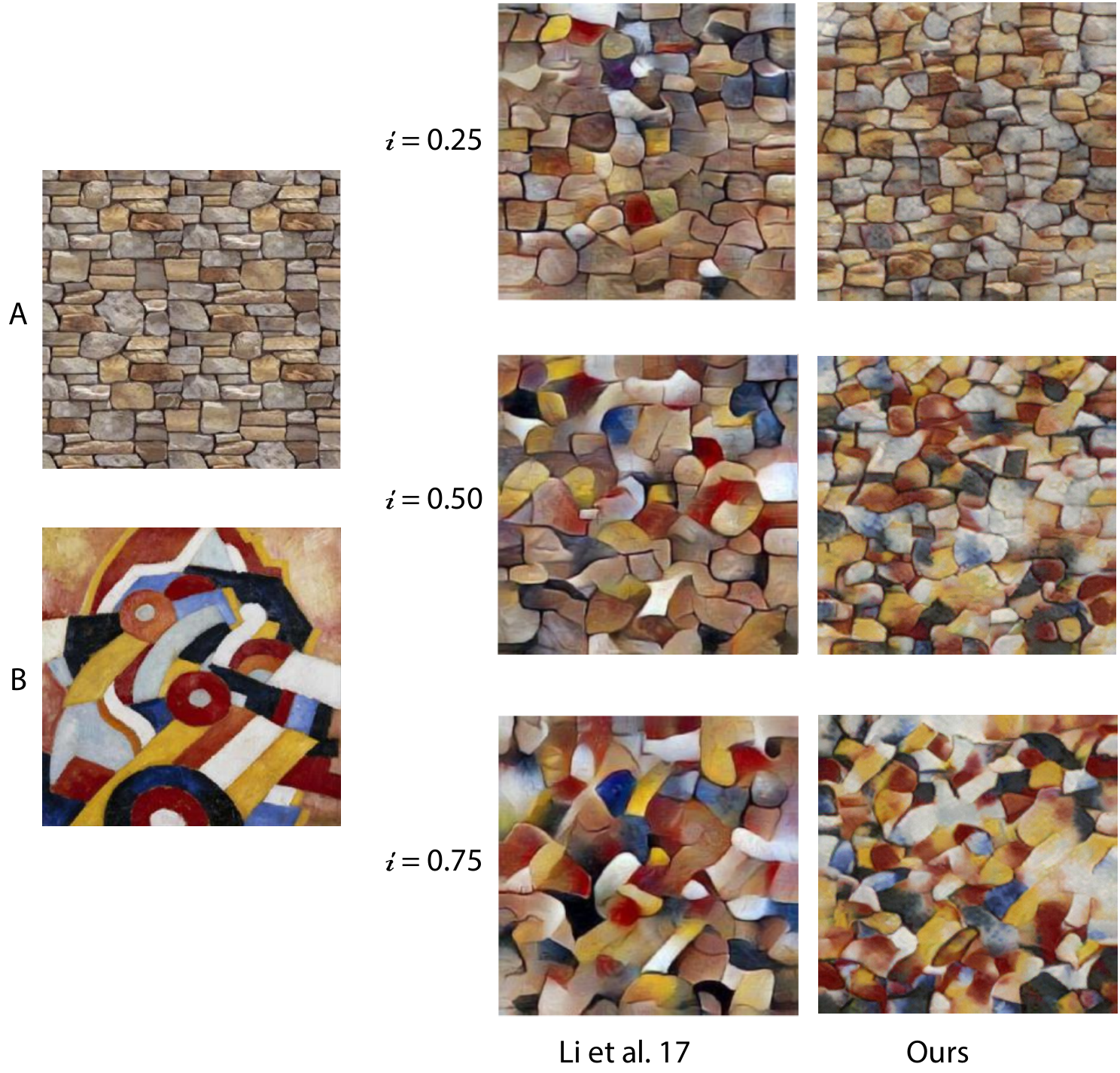}
	\caption{Comparisons of texture mixing against Li et al. 2017, we achieve a more consistent homogeneous mixture where deep features appear to interpolate rather than fight for local dominance. } 
	\label{fig:mixing2}
\end{figure}

\subsection{User Controls}

\begin{figure}[hbt!]
	\centering
	\setlength{\h}{3in}
	\includegraphics[height=\h]{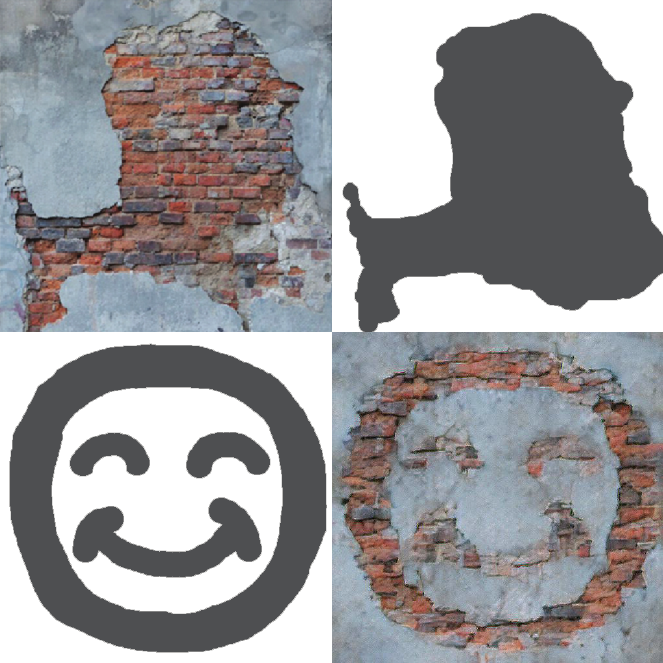}
	\caption{Top row: input image and corresponding style mask. Bottom row: New user defined target "content" mask and corresponding synthesized output.} 
	\label{fig:painting}
\end{figure}

\begin{figure*}
	\centering
	\includegraphics[width=17.5cm]{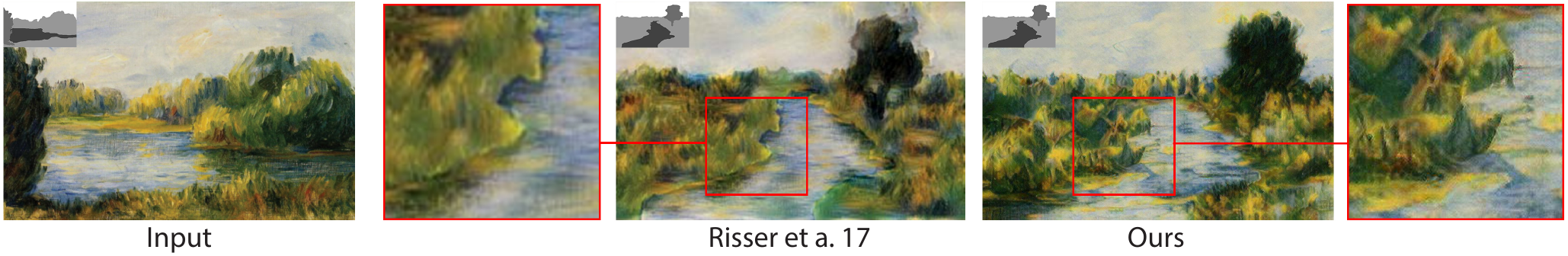}
	\caption{Comparison against the previous painting-by-numbers approach that utilizes multiple parametric models. We notice superior results at the border regions between different textures.} 
	\label{fig:paintingComparison}
\end{figure*}

The utility of texture synthesis as an artistic tool is marked by how easily and intuitively a user can guide the process and control the final output. Previous user controls for texture synthesis methods typically employ the "painting-by-numbers" strategy, where discrete masks are used to divide-and-conquer the global synthesis operation into a collage of local and independent parametric models, one for each associated texture ID in the mask. This is effective for large coherent regions, but is marked by poor transition areas between textures, along with an efficiency cost associated with each additional texture ID. In contrast, optimal transport can be guided simply by re-balancing and re-weighting the feature space statistics.

Given masks that assign a texture ID to each pixel in the content and style images, there are two modifications added to the core algorithm. First, the target PDF $S$ must be re-balanced so that its feature histogram with respect to texture IDs matches the desired histogram for the content mask. This can be achieved by simply  removing or duplicating samples from each histogram bin at random. During synthesis $O$ requires an additional processing step so that image regions with a given texture ID are more likely to map to similar texture ID regions of $S$. We found that a naive approach can achieve satisfactory results. Before the optimal transport operation, re-weight the the distribution for each content histogram bin so that the distributions mean matches the distribution mean of the corresponding bin in the target histogram. While this is a relatively loose constraint, it appears to sufficiently bias the optimal transport operation so that features are anchored to the desired image locations while allowing for the transition areas between texture regions enough flexibility to map to their optimal region of the the target PDF $S$. 

This is illustrated in figure \ref{fig:painting} where a simple heterogeneous texture containing two continuous homogeneous sub-textures are masked and re-targeted using simple re-balancing and re-weighting. We see that the content mask sufficiently guides the synthesis process of coherent homogeneous regions while allowing the optimal transport process enough flexibility to reproduce the novel and complex transition features between regions, shown here as peeling and cracking plaster. This is in direct contrast to previous divide-and-conquer methods that utilize multiple parametric models \cite{Risser17} as shown in figure \ref{fig:paintingComparison}. This illustration shows a more complex image comprising more distinct homogeneous textures that have more sophisticated and incomplete transitions between regions. We highlight one such synthesis border region where bush features are re-targeted on the left and river features on the right. No corresponding border feature exists in the input and we see the previous parametric approach \cite{Risser17} struggles with this problem, producing a sharp and discontinuous seam between the masked regions. Our method however is able to achieve more natural transitions both in ambiguous regions as well as border regions in general.

This section has highlighted the texture painting use case because it most directly illustrates the power of statistics re-balancing and re-weighting as a simple means of guiding the optimal transport process. It should be noted that this approach can and should also be used to guide the style transfer and texture mixing process as well. The results shown in this section use the full texture synthesis and style transfer algorithm with the content strength set to zero and the starting image set to noise.

\section{Quality}

We now discuss some advantages of our results. By ensuring that the full statistical distribution of exemplar features are preserved, our optimal transport approach addresses the instabilities commonly observed in neural texture synthesis methods that utilize parametric texture models or other summary statistics. While adding a histogram loss ~\cite{Risser17} to the parametric model can also fix instabilities, this approach is more complicated and requires multiple loss functions that are difficult to keep in balance. It also relies on the use of back-propagation training, making it too slow and impractical for real-world usage. We compare our optimal transport optimization results against the WCT approach ~\cite{Li17DiversifiedTextureSynthesis} because both share the same auto-encoder synthesis strategy and because the WCT transform behaves as a proxy for the Gram/Covariance matrix texture models commonly used in the related literature.  

We find that our optimal transport approach outperforms the WCT strategy in multiple ways: 
\begin{enumerate}
\item Larger structured features in the texture/style are better represented by the first-order joint statistics of the full feature distribution.
\item Feature blending/smearing artifacts of WCT are significantly reduced by our approach due to the additional "slices" capturing a more detailed view of the feature distribution.
\item Content and style features are optimized together, rather than competing as discussed in section \ref{sec:style}.   
\item Our optimal transport framework unifies style and color, a known open problem.
\item Mixing textures produces a more homogeneous result with more convincing interpolations of individual features.
\item A simple re-balancing and re-weighting strategy allows users to guide both the texture synthesis and style transfer process.
\end{enumerate}

\begin{figure*}
	\centering
	\includegraphics[width=17cm]{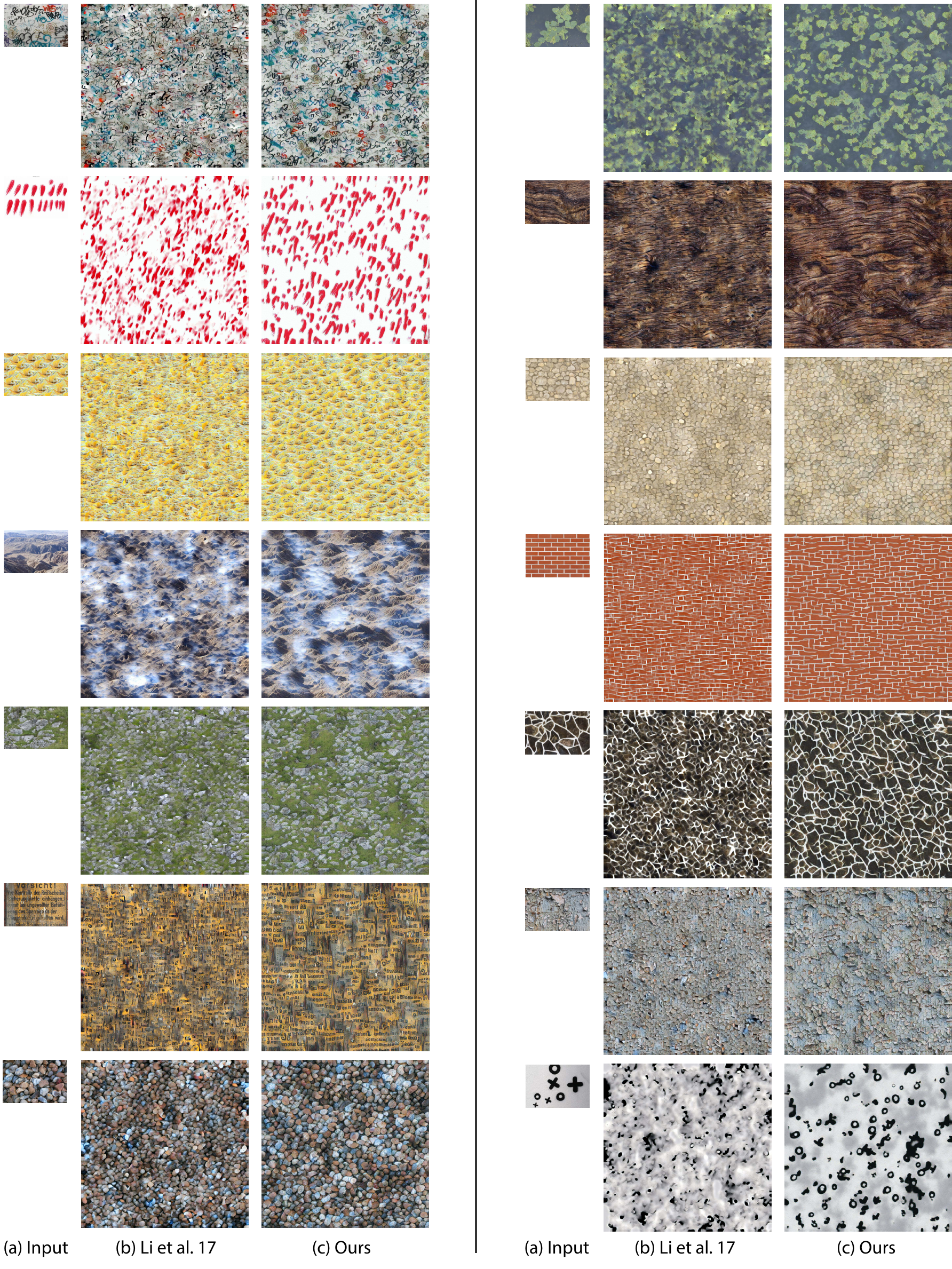}
	\caption{Results of the (b) WCT of Li et al. 2017 in comparison to (c) our Optimal Transport approach.} 
	\label{fig:textureResults}
\end{figure*}

\begin{figure*}
	\centering
	\includegraphics[width=17cm]{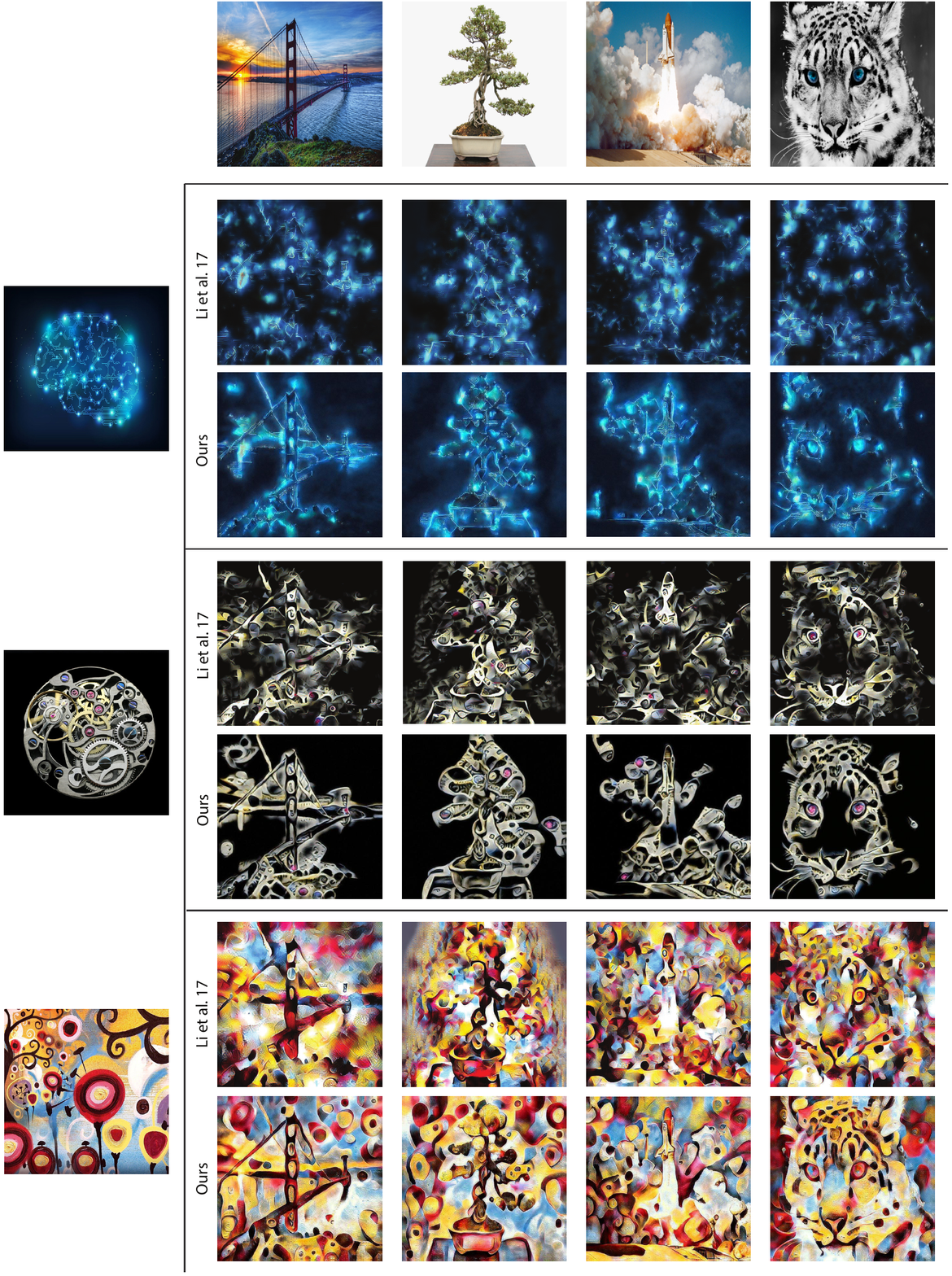}
	\caption{Results of our method for style transfer compared with the WCT method of Li et al. 2017.} 
	\label{fig:styleTransferResults}
\end{figure*}

Side-by-side comparisons between the WCT and our Optimal Transport method are provided in figures \ref{fig:textureResults} and \ref{fig:styleTransferResults}. 

\textbf{Speed}, is a key benefit of our algorithm. Running times for our method are as follows. We used a machine with four physical cores (Intel Core i5-6600k), with 3.5 GHz, 64 GB of RAM, and an Nvidia Quadro P6000 GPU with 24 GB of GPU RAM, running Ubuntu. For a single 1024x1024 image, our method takes 23 seconds utilizing PCA and 84 seconds without PCA. This is in contrast to the back-propagation based optimization methods such as Risser~et~al. \shortcite{Risser17} and Gatys~et~al. \shortcite{gatys2016image} that takes tens of minutes. Our approach used three pyramid levels. For style transfer we add a progressively weighted content matching at relu3\_1, relu4\_1 and relu5\_1 which increases the running time by a negligible amount. These metrics were measured over 100 full image synthesis operations. We believe this run-time performance makes optimal transport an attractive candidate for an interactive artist tool, particularly when only sub-regions of the image are edited in real time. Our current implementation utilizes a mixture of CPU and GPU processing, incurring a large performance penalty when synchronizing memory. We believe that significant performance improvements could be achieved through a strict GPU implementation.

\section{Conclusion}

We believe that directly matching feature statistics is the native problem formulation for Texture Synthesis and by doing so, we are able to use optimal transport to achieve unprecedented speed and quality for texture synthesis while also solving a wide range of Texture Synthesis-based problems that were previously believed to require separate techniques or non-trivial extensions to the core algorithm. We propose a simple, well-principled method for Texture Synthesis and its many sub-fields: Style Transfer, Texture Mixing, Inverse Texture Synthesis and Texture Painting. We present N-Dimensional probability density function transformations through an iterative sliced histogram-matching operation as the core component to a truly universal and general purpose texture synthesis algorithm.

\textbf{Future Work} We believe our approach can lay the groundwork for several interesting future research directions. We believe that synthesis quality could be further improved through both the study of superior encoding networks that are designed and trained specifically for texture recognition as well as methods for training a more accurate decoder network. We view the image degradation resulting from VGG encoding followed by decoding as the key shortcoming of ours and previous methods that rely on the VGG auto-encoder framework \cite{Li17DiversifiedTextureSynthesis} and warrants further exploration. This paper was largely inspired by histogram-guided texture synthesis along with a body of color transfer literature. We believe this paper serves as a bridge connecting the two fields and opens the way for future cross-pollination of these theoretically related yet historically separate topics.

\textbf{Acknowledgments} We would like to acknowledge and thank Keyang Xiang for his assistance with implementation and Akash Garg, David Harmon and Marc Ellens for their peer reviews and general suggestions for improving this paper.



\bibliographystyle{acmsiggraph}
\nocite{*}
\bibliography{ms}
\end{document}